\documentclass[final]{aipproc}

\layoutstyle{8x11single}

\def\lsim{\,\lower2truept\hbox{${< \atop\hbox{\raise4truept\hbox{$\sim$}}}$}\,}
\def\gsim{\,\lower2truept\hbox{${> \atop\hbox{\raise4truept\hbox{$\sim$}}}$}\,}

\SetInternalRegister\hbadness{8000} % pseudo latin isn't breaking very well :-)

\newcommand\doingARLO[2][]{%
  \ifx\mmref\undefined #1\else #2\fi
}

\begin{document}

\title 
      [Polarized Galactic radio emission at degree scales]
      {A Multifrequency Analysis of the Polarized Diffuse Galactic Radio 
Emission at Degree Scales}


\author{C. Burigana}{
  address={Istituto TeSRE/CNR, via Gobetti 101, I-40129 Bologna, Italy},
  email={burigana@tesre.bo.cnr.it},
  thanks={This work was commissioned by the AIP}
}

\iftrue
\author{L. La Porta}{
  address={Istituto TeSRE/CNR, via Gobetti 101, I-40129 Bologna, Italy},
  email={laporta@tesre.bo.cnr.it},
}

\fi

% \copyrightholder{Acoustical Scociety of America}
\copyrightyear  {2001}

\begin{abstract}
The polarized diffuse Galactic radio emission, mainly synchrotron emission,
is expected to be one of the most relevant source of astrophysical contamination
at low and moderate multipoles in cosmic microwave background 
polarization anisotropy experiments 
at frequencies $\nu \le 50 \div 100$~GHz.  
We present here preliminary results based on a recent analysis of the 
Leiden surveys covering about 50\% of the sky at low
as well as at middle and high Galactic latitudes. 
By implementing specific interpolation methods to 
deal with these data, which show a large variation of the sampling across
the sky, we produce maps of the polarized diffuse Galactic synchrotron 
component at frequencies between 408 and 1411~MHz with pixel sizes
larger or equal to $\simeq 0.92^\circ$. We derive the angular power spectrum
%(in terms of antenna temperature), $C_\ell^{ant}$,
of this component for the whole covered region and for three patches 
in the sky significantly oversampled with respect to the average and 
at different Galactic latitudes. We find multipole spectral 
indices typically ranging between $\sim -3$ and $\sim -1 \div -1.5$,
according to the considered frequency and sky region.
At $\nu \ge 610$~MHz, the frequency spectral indices observed in the considered
sky regions are about $-3.5$, compatible with an intrinsic frequency 
spectral index of about $-5.8$ and a depolarization due to Faraday
rotation with a rotation measure RM of about $15$~rad/m$^2$.
This implies that the observed angular power spectrum of the polarized
signal is about 85\% or 20\% of the intrinsic one
at 1411~MHz or 820~MHz respectively.
\end{abstract}

\date{\today}

\maketitle

\section{Introduction}

The polarized diffuse Galactic synchrotron emission,
whose study provides important insights into the properties of
the Galactic magnetic field and the interstellar ionized matter,
is expected to be one of the most relevant source of astrophysical contamination
at low and moderate multipoles in 
cosmic microwave background (CMB) polarization anisotropy experiments 
at frequencies $\nu \le 50 \div 100$~GHz. At frequencies about 1~GHz, the Galactic 
synchrotron emission dominates over the Bremsstrahlung emission
which, although expected to be weakly polarized, significantly increases
with the frequency, $\nu$, in comparison with the synchrotron one.

While ongoing and future experiments with 
high sensitivity and resolution are expected to cover large sky areas 
at different Galactic latitudes and frequencies (see, e.g., these Proceedings),
the Leiden surveys [1], covering about 50\% of the sky, can be used 
to derive the angular power spectrum of the polarized diffuse Galactic radio 
emission at frequencies between 408 and 1411~MHz.
We have implemented specific interpolation methods 
to project surveys with large variations of the 
sampling across the sky into maps;
we work here with pixel sizes larger or equal to $\simeq 0.92^\circ$,
appropriate to the case of the Leiden surveys.
Some sky areas, both at low and middle/high Galactic latitudes, 
show a much better sky sampling than the average 
and are particularly suitable for an analysis in terms of angular
power spectrum. Their extents, few tens of degrees both in Galactic latitude
$b$ and longitude $l$, together with the beamwidths 
(HPBW from $2.3^\circ$ to $0.6^\circ$ for $\nu$ from 408 to 1411~MHz) 
and the measure sensitivities (of about 100~mK at 610~GHz and 
about 1.5 (3) times better (worst) at the highest (lowest) frequency)
imply that, in principle,  we can study 
the angular power spectrum, $C_\ell^{ant}$ 
(here in terms of antenna temperature),
in a multipole range from $\ell \sim$~few~tens, where boundary effects 
begin to be neglibible, to about $\ell \sim 100$, where, even for the 
better sampled regions, the noise power is close to the signal power.
In addition, the survey full coverage analysis allows to recover
the $C_\ell^{ant}$s at $\ell$ between $\sim 5$ and few tens.  

As a by-product of this work, these maps of polarized diffuse Galactic 
emission, once properly 
rescaled in frequency, 
can be used as inputs for simulation activities in current and 
future microwave polarization anisotropy experiments.

\section{Map production and consistency tests}

The main problem in the analysis of the Leiden 
surveys is due to their poor sampling across the sky; it is necessary 
to project them into maps with pixel size of about $3.7^\circ$
(i.e. $n_{side}=16$; we use here the HEALPix scheme [2]
in which the number of pixel in the sky is $12 n_{side}^2$), 
to find about one observation into each pixel of the whole 
observed sky region. 
For the same reason, a smoothing of the whole data with a 
window function with comparable size can not be properly applied.
On the other hand, for some sky regions the sampling of the surveys
is significantly much better, by a factor $\simeq 4$. We identified
three regions (see [3]):
patch 1 [($110^\circ \le l \le 160^\circ$, $0^\circ \le b \le 20^\circ$)];
patch 2 [($5^\circ \le l \le 80^\circ$, $b \ge 50^\circ$) together with
($0^\circ \le l \le 5^\circ$, $b \ge 60^\circ$) and
($335^\circ \le l \le 360^\circ$, $b \ge 60^\circ$)];
patch 3 [($10^\circ \le l \le 80^\circ$, $b \ge 70^\circ$)].

Therefore, we chose to derive the angular power spectrum 
at $\ell$ between $\simeq 5$ and $\simeq 50$, 
as representative of the ``whole'' sky, 
by working with the survey full coverage, 
and at $\ell$ between $\simeq 30$ and $\simeq 100$ by working 
with the above patches, which allow to analyze both low and
middle/high Galactic latitudes.
We anticipate that in these patches the polarized signal is typically higher 
than the average and it also shows relevant intensity
variations on scales of some degrees. We then expect to find in the patches
an angular power spectrum relatively higher than the spectrum obtained from the 
survey full coverage analysis in the common multipole range.

By simply averaging the survey observations in a given pixel, 
we can produce maps of polarized signal $P$ (and also of Stokes parameters
$Q$ and $U$) that can be analyzed in terms of $C_\ell^{ant}$s
of the polarized signal (and $E$ and $B$ modes).
On the other hand, we verified that the results produced in this way,
although in rather good agreement with those derived from the improved method
described below both in terms of maps and of $C_\ell^{ant}$s,
are affected by discontinuity effects on 
scales of the order of the pixel size. 
This tends to add power in the recovered $C_\ell^{ant}$s,
particularly at multipoles close to that corresponding to 
the pixel size [$\ell \sim 180/(\theta_{\rm pix}/{\rm deg})$], 
because of its ``euristic'' similarity with point source confusion noise.

To overcome this problem, we implemented a specific ``interpolation'' 
method and decided to produce maps at resolution of about $0.92^\circ$
(in order to smooth the discontinuities discussed above on scales
of about $2^\circ$), which, of course, provide reliable information
only on scales larger than $\simeq 2^\circ$, i.e. only up to $\ell \simeq 100$.
We assign to each sky pixel the average of the signals
falling close to the pixel centre, properly weighted according to
a certain power of the distance from the pixel. Different powers 
have been tested (from $\sim 0.5$ to $\sim 2$): clearly, the higher
the power the higher the map contrast. On the other hand,
we verified that the results depend very weakly on the adopted power.
The algorithm searches, pixel by pixel, for a suitable 
number of observations to use in the weighted average, according to the 
following basic recipes: 
(i) to have enough observations (typically more than 3, if compatible
with the following criteria); (ii) to use only observations quite close to the
considered pixel centre (typically, less than few degrees, if compatible with 
the other criteria); (iii) to obtain the convergency of the result 
(i.e. minimize its fractional variation) with the variation of the 
number of observations and of the circle around the pixel centre 
that contain them.
By applying our algorithm with different resolutions (we consider here
$n_{side}$ from 16 to 64), we can test its dependence on the details
of its practical 
application\footnote{We tested also a mix of a simple average 
of the signals in the pixel - whenever possible - and of 
this ``interpolation'' scheme. We found preferable to apply everywhere
the ``interpolation'' scheme, in order to reduce better the 
discontinuity effects.}.
Clearly, each interpolation method may alter the power 
at small scales; in principle, it operates as a kind of filter or 
regularization of the map and may decrease the power at small scales. 
Then, a reasonable agreement with the angular power spectrum 
derived, whenever possible, from other data with a better sampling 
across the sky (at the same $\nu$ and in same sky region) 
is crucial to probe the validity of the code recipes: in particular,
for the reason discussed above, we require that it does not 
produce an underestimation of the power spectrum.

We obtained maps for $P$, $Q$, 
and $U$ at all frequencies and with 
different resolutions ($16 \le n_{side} \le 64$) by using 
the different methods discussed above
\footnote{We verified also, pixel by pixel in the maps, that
$P \simeq (Q^2+U^2)^{0.5}$ with an accuracy significantly better 
than the rms noise in each pixel.}.

From each map,
by using the HEALPix package (properly the ``anafast'' code)
we derive the angular power spectrum of the polarized signal
for the survey full coverage and for the three considered patches 
(we renormalized the $C_\ell^{ant}$s to the case of whole sky coverage).
Preliminary results at 1411~MHz 
as well as tests of the applicability of this method to derive 
the $C_\ell^{ant}$s on relatively small patches have been reported in [3]
(in our case the limited patch extent is a minor
problem, because of the larger dimension of the considered sky areas).

In the next section we present some of our results and show how 
both the maps and the angular power spectra derived from them pass
all the consistency tests discussed above.

\section{Results}

In Fig.~1 we show one of the maps of polarized signal produced 
with the code described in the previous 
section\footnote{A color figure may be requested via e-mail to the authors.}.

In Fig.~2 we compare the angular power spectra of the patch 1 at 610~MHz 
obtained by working at different
resolutions: note the good agreement in the common multipole range. The 
same holds at all frequencies for the different sky regions considered here
and also by extending the comparison to the lowest resolution maps 
($n_{side}=16$), a crucial test in the case of the 
survey full coverage analysis.

The angular power spectrum derived from each map is, of course, mainly
given by the sum of the astrophysical diffuse component relevant here,
the synchrotron, and of the instrumental noise; in particular, the latter
dominates at large multipoles, as it is evident also from the flattening 
we find there.
We fit the recovered $C_\ell^{ant}$s as sum of two components, 
represented by a set of parameters:
(i) the synchrotron component, 
smoothed with the beam (assumed to be perfectly symmetric and Gaussian), 
is approximated 
as $k \ell^\alpha {\rm exp}[-(\sigma_b \ell)^2]$, $\sigma_b$ being
the 1$\sigma$ beamwidth (of course, the window function is not crucial
in this context at the highest frequencies);
(ii) the noise contribution is approximated
as a flat, white noise, component, $c_{wn}$. All the parameters of the fit 
depend on $\nu$ and on the sky region. In each case, we find the best
fit parameters at $\ell$ between $\sim 30$ and $\sim 100 \div 200$;
we separately repeat the fit at $\ell$ between $\sim 5$ and $\sim 30$ 
in the case of the survey full coverage analysis.

\begin{figure}
  \caption{Map of polarized signal obtained at 
$n_{side}=64$ ($\simeq 0.92^\circ$ of pixel size).}
\includegraphics[height=.5\textheight,angle=90]{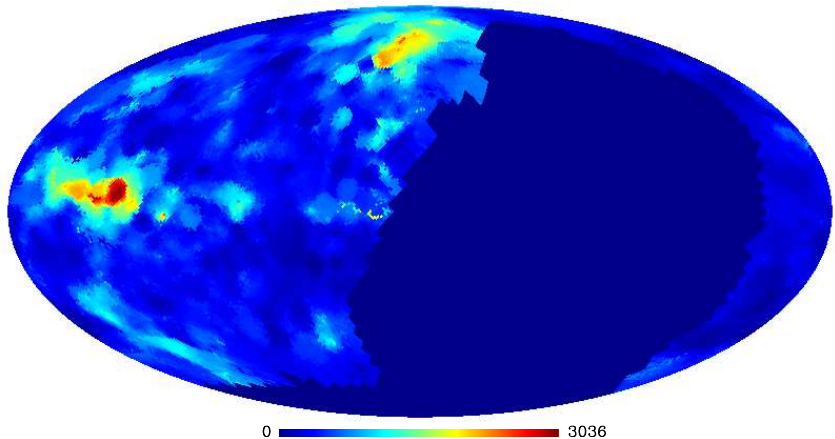}
\end{figure}

\begin{figure}
  \caption{Comparison between the angular power spectra obtained by projecting
the Leiden surveys into maps at different resolution, 
$n_{side}=32$ (i.e., pixel size $\simeq 1.8^\circ$,
dotted line) and 64 (i.e., pixel size $\simeq 0.92^\circ$, solid line). 
%Note the good
%agreement between the two angular power spectra up to 
%$\sim$ the maximum multipoles reachable by analyzing 
%the lowest resolution map.
}
\includegraphics[height=.7\textheight,width=.32\textwidth,angle=90]{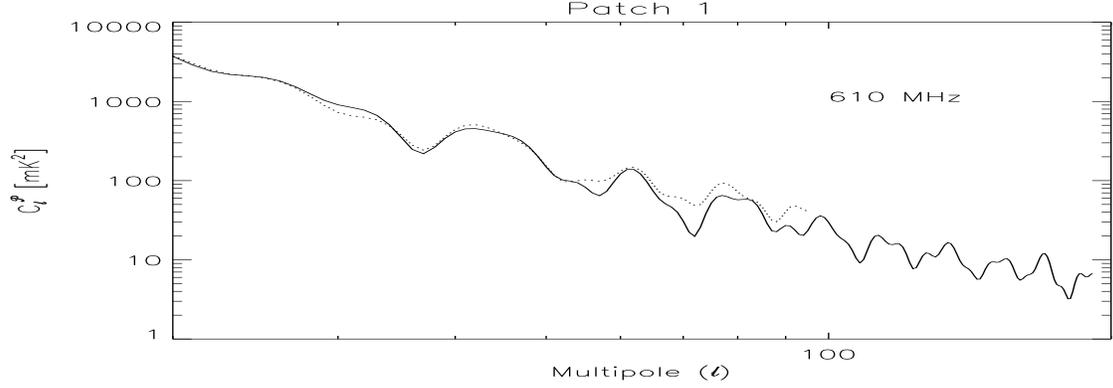}
\end{figure}

Of course, it is  extremely important to check that the level of the noise
power derived from the fit is similar to that derived on the basis of
the rms noise per pixel in the map. 
Given rms noise maps, we generate simulated maps of white noise
and extract their $C_\ell^{ant}$s.
This is shown in Fig.~3.
We quote the rms noise per pixel by using three different methods:
(i) the rms noise calculated by using the rms noise of each observation
and propagating the error according to the 
mathematical rules applied to produce the maps (solid lines);
(ii) the same as in the case (i), except for pixels in which survey 
observations fall, in which case we apply the standard weighted error on the 
standard weigthed average (dot-dashed line in the left panel);
(iii) the standard weighted error on the 
standard weigthed average for pixels in which survey
observations fall, and the average of these errors for the other pixels
(three dots-dashes). Of course, for the patches, and particularly for the patch 3,
the evaluations (ii) and (iii) provide practically the same results, given the
good sampling across the sky; thus, we only report the evaluation (iii).
Note that, for the survey full coverage analysis, 
the noise spectrum evaluated according
to (iii) is of the order of that derived from the signal map
at $\ell \sim 50$: 
therefore, at $\ell$ larger than $\sim 50$ 
no reliable information can be derived.
For the patches, on the contrary, the noise spectrum is below
that derived from the signal map, independently of the different 
noise evaluations, up to $\ell \sim 100$. This confirms that,
not only from the point of view of the sampling across the sky, but also
from the point of view of the sensitivity, the patches can be used to 
extract the synchrotron $C_\ell^{ant}$s up to $\ell \sim 100$. 
We note that this is not valid, in principle, for possible analyses of 
patches in other sky regions less sampled 
and of lower polarized signal intensity,
where we expect a signal 
to noise ratio close to that derived here for the survey full coverage 
analysis,
i.e. a reliable information can be obtained only up to $\ell \sim 50$. 

From the survey full coverage analysis, 
we derive reliable information also at $5 \le \ell \le 30$
(of course, only nearly full sky observation can provide reliable information 
at very low multipoles). In this case, both the noise and possible 
effects introduced by the poor sampling or by the adopted ``interpolation'' 
technique are clearly negligible.
We find a flattening of the spectrum with respect to that at higher multipoles;
the slope (dotted line of left panel of Fig.~3) 
at $\nu = 1411$~MHz, crucial for the extrapolation to higher 
frequencies, is close to $-2.5$. This implies an essentially 
flat (slope $\sim -0.25$) behaviour of 
$\delta T_\ell = \sqrt{\ell(2\ell+1)C_\ell/4\pi}$ at $5 \le \ell \le 30$
($\delta T_\ell$ is a quantity particularly 
relevant in the comparison with the CMB temperature and polarization anisotropy 
observations). 

\begin{figure}
  \caption{Angular power spectra derived from the map (thin solid lines) and
synchrotron (thick solid lines) and synchrotron plus noise (thick 
dot-dashed lines) best fit spectra at multipoles larger than $\sim 30$
compared with the those derived from simulated maps of pure white noise
with different estimates of the rms noise per pixel (roughly horizontal lines).
In the case of the survey full coverage analysis we report 
also the $C_\ell^{ant}$s
obtained at lower multipoles (again, thin solid line)
together with the best fit spectrum (dotted line) dominated
by the synchrotron component (see also the text).}
\includegraphics[height=.25\textheight,width=1.\textwidth]{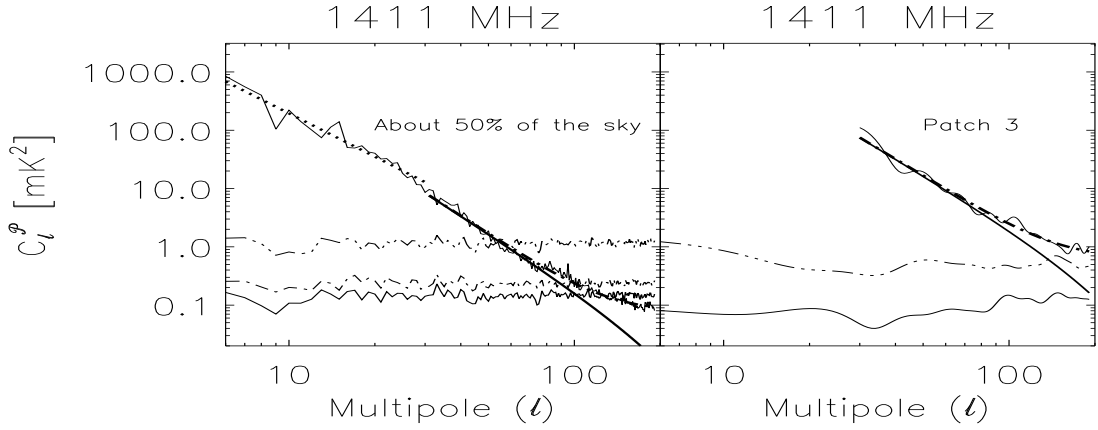}
\end{figure}

We focus now on the angular power spectrum 
of the synchtrotron component at $\ell$ between $\simeq 30$ and 
$\simeq 100 \div 200$.
Our results are summarized in Fig.~4. 
We find slopes varying from about $-3$ to about $-1 \div -1.5$,
mainly depending on the frequency, although, as expected, 
different sky regions show different slopes even at the same frequency.

As already recognized (see [3]), we find 
a particularly impressive agreement 
between the result obtained for the patch 1 and 
the angular power spectrum (long dashes in Fig.~4) obtained 
at 1411~MHz in a smaller region (the region 2 in eq. (20)
of [3]) inside the patch 1 by exploiting 
the data from [4] (which have
better sampling across the sky, sensitivity and resolution and
in which only the absolute calibration takes
advantage of the Leiden surveys).
This strongly supports the validity of our code to project 
the Leiden surveys into HEALPix maps or, at least, indicates 
that it does not introduce significant errors in 
the evaluations of the $C_\ell^{ant}$s
even at $\ell \sim 100$ and in the worst case of 
$\nu = 1411$~MHz (here the ratio between the beamwidth 
and the typical angular distance among adjacent survey pointings 
is minimal and the survey sky coverage is not the best 
-- it occurs at 610~MHz).

\begin{figure}
  \caption{Synchrotron component of the angular power spectrum derived 
for the entire map and the three patches at the different frequencies:
408 (solid lines), 465 (dotted lines), 610 (dashed lines), 820 
(dot-dashed lines) and 1411~MHz (three dots-dashes). In the case
of the patch 1, note the good agreement (within a factor $\sim 2$)
with the spectrum (long dashes) obtained from a smaller patch 
inside the patch 1 
by exploiting the survey [4] at 1411~MHz (see also the text).}
\includegraphics[height=.7\textheight,width=.5\textwidth,angle=90]{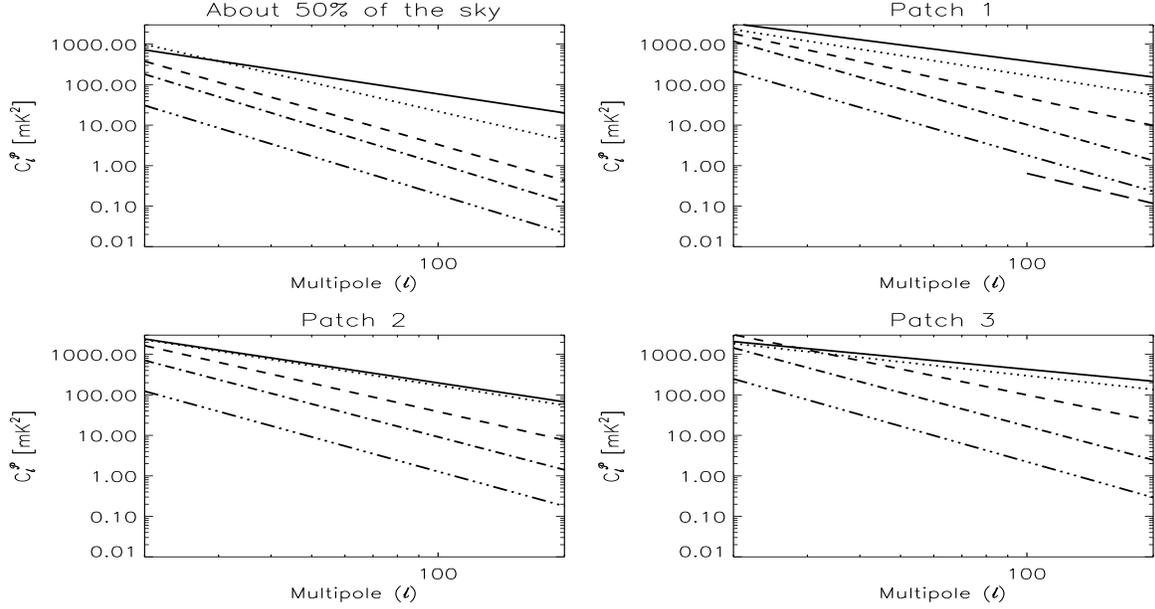}
\end{figure}

\section{Discussion and Conclusions}

We presented here preliminary results based on a recent analysis of the 
Leiden surveys for about 50\% of the sky and 
three patches at low
as well as at middle/high Galactic latitudes
significantly better sampled than the average.
By implementing specific interpolation methods, 
we produce maps of the polarized diffuse Galactic synchrotron 
component at frequencies between 408 and 1411~MHz with pixel sizes
larger or equal to $\simeq 0.92^\circ$. 

%As a by-product, these maps can constitute, when properly 
%rescaled in frequency, inputs for simulation activities in current and 
%future microwave polarization anisotropy experiments.

We derive the angular power spectrum
of the polarized diffuse Galactic synchrotron 
emission for the whole covered region and for the three patches. 
The multipole spectral indices typically range between $\sim -3$ and 
$\sim -1 \div -1.5$,
depending on the frequency and on the sky region.

Together with the large sky coverage, the multifrequency 
Leiden surveys offer the opportunity
to study the dependence of the Galactic synchrotron polarized signal on $\nu$.
Clearly, depolarization effects and possible transitions 
from essentially optically thin regimes to significantly self-absorbed 
ones may ``obscure'' the intrinsic synchrotron spectral behaviour.
We find large variations of the spectral index observed in the different sky 
pixels, probably due to real variations, but also to 
the limited sensitivity and, perhaps, to differential depolarization 
effects related to the $\nu$-dependent beamwidth or, 
finally, to possible systematic and not well understood effects in the data. 
We try to circumvent, at least in part, 
these difficulties by exploiting the 
``statistical'' information contained in the $C_\ell^{ant}$s. 

In Fig.~5 we show the angular power spectrum as a function of $\nu$
at the multipole $\ell =50$ (a middle, representative 
value considered in this study); we find very similar results
also at $\ell = 30$ and 100. 
The observed frequency spectral indices 
at $\ell =50$ are $\sim$ $-3.2$ (survey full coverage analysis),
$-3.3$ (patch 1), $-3.6$ (patch 2) and $-3.9$ (patch 3).
At the highest frequencies (820, 1411~MHz), 
that are crucial for the extrapolation
at frequencies of $20\div100$~GHz relevant for CMB anisotropy 
polarization radiometric experiments 
and where beamwidth depolarization effects are expected to be
quite similar (HPBW varying only of $\sim 50$\% between 820 and 1411~MHz), 
the observed slope can be easily explained by assuming an 
intrinsic slope of $-5.8$ (as in the case of a slope $-0.9$ in terms of flux)
and a depolarization due to Faraday rotation with a rotation 
measure RM of about $15$~rad/m$^2$.
This results is in quite good agreement with that previously obtained 
from the analysis of a sky region partially overlapped with the patch 1, 
as well as clearly compatible with the upper limits on RM 
found in the Leiden surveys (see [5]).
This implies that the observed angular power spectrum of the polarized
signal is about 85\% or 20\% of the intrinsic one
(strictly, in absence of Faraday rotation depolarization) 
at 1411~MHz or 820~MHz respectively.

\begin{figure}
  \caption{Angular power spectrum at multipole $\ell = 50$ as function
of $\nu$ for the different sky regions:
the survey full coverage (solid lines), the patch 1 (dotted lines), 
the patch 2 (dashed lines) and the patch 3 (dot-dashed lines).
The straight longer lines represent a simple power law dependence of the
$C_\ell^{ant}$s (at $\ell = 50$) on $\nu$ corresponding to 
an ``intrinsic'' slope of $-5.8$, 
rescaled to the angular power spectra
derived here at 1411~MHz. The straight shorter lines 
represent the dependence of the $C_\ell^{ant}$s 
on $\nu$ between 820 and 1411~MHz 
for a Faraday rotatation depolarization
with RM~=~15~rad~/~m$^2$ applied to the above ``intrinsic'' slope
(in the case of the survey full coverage analysis, note 
the practically perfect agreement with the observed spectrum
between 820 and 1411~MHz).
Note also the further flattening 
of the spectrum at the lowest frequencies in the case of the two
patches at relatively high Galactic latitudes (see also the text).} 
\includegraphics[height=.7\textheight,width=.45\textwidth,angle=90]{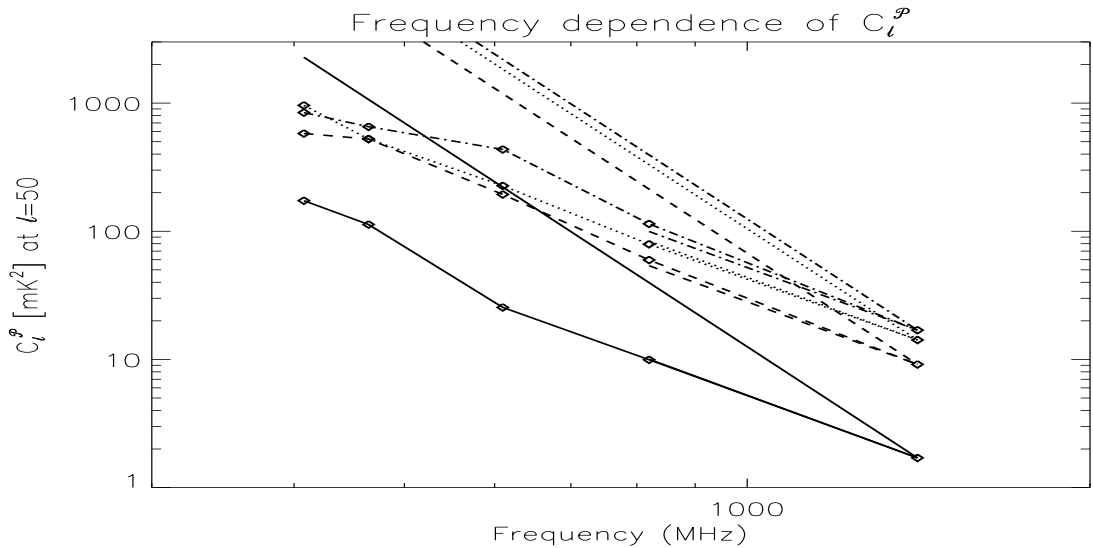}
\end{figure}

\begin{theacknowledgments}
It is a pleasure to thank C.~Baccigalupi, G.~De~Zotti and R.~Fanti
for many constructive comments. 
We acknowledge also D.~Maino, M.~Maris, R.~Paladini and F.~Perrotta 
for the fruitful collaboration on this topic.
We are grateful to T.A.T. Spoelstra for his kind clearifications and to
L.~Chiappetti and P.~Platania for useful explanations 
on the database of the Leiden surveys, provided us in a more 
user-friendly form. The HEALPix package is acknowledged.
\end{theacknowledgments}

\section{References}

1. Brouw W.N., Spoelstra T.A.T., A\&AS, {\bf 26}, 129 (1976)

\noindent
2. G\`orski K.M., Hivon E., Wandelt B.D. 1998, 
``Analysis Issues for Large CMB Data Sets'',
in {\it MPA/ESO
Conference on Evolution of Large-Scale 
Structure: from Recombination to 
Garching}, edited by A.J. Banday, R.K. Sheth, L. Da Costa, 37, astro-ph/9812350

\noindent
3. Baccigalupi C., Burigana C., Perrotta F., et al., A\&A, {\bf 372}, 8 (2001)

\noindent
4. Uyaniker B., Furst E., Reich W., Reich P., Wielebinski R., 
A\&AS, {\bf 138}, 31 (1999)

\noindent
5. Spoelstra T.A.T., A\&A, {\bf 135}, 238 (1984)

% choose bibtex style depending on layout style and options used in
% sample:

\doingARLO[\bibliographystyle{aipproc}]
          {\ifthenelse{\equal{\AIPcitestyleselect}{num}}
             {\bibliographystyle{arlonum}}
             {\bibliographystyle{arlobib}}
          }
\bibliography{sample}

\end{document}